\documentclass[conference]{IEEEtran}
\IEEEoverridecommandlockouts
\usepackage{amsmath,amssymb,amsfonts}

\usepackage{algorithmic}
\usepackage{graphicx}
\usepackage{textcomp}
\usepackage{xcolor}
\usepackage{amsthm}
\usepackage[parfill]{parskip}
\makeatletter
\newcommand{\removelatexerror}{\let\@latex@error\@gobble}
\makeatother
\usepackage{float}
\usepackage[portuguese,lined,boxed,ruled]{algorithm2e}
\usepackage[english]{babel}
\usepackage[utf8]{inputenc}
\usepackage[numbers]{natbib}
\usepackage{multicol}
\usepackage{multirow}
\everymath{\displaystyle}
\def\BibTeX{{\rm B\kern-.05em{\sc i\kern-.025em b}\kern-.08em
    T\kern-.1667em\lower.7ex\hbox{E}\kern-.125emX}}

\addto\captionsbrazil{
\renewcommand{\figurename}{Fig.}
}

\begin{document}

\title{2D Basement Relief Inversion using Sparse 
 Regularization.\\
\thanks{}
}

\author{
\IEEEauthorblockN{Francisco Márcio Barboza}
\IEEEauthorblockA{\textit{Department of Computing and Technology} \\
\textit{Federal University of Rio Grande do Norte}\\
Caicó, Rio Grande do Norte \\
marcio.barboza@ufrn.br}
\and
\IEEEauthorblockN{Arthur Anthony da Cunha Romão E Silva}
\IEEEauthorblockA{\textit{Graduate Program on Systems and Computing} \\
\textit{Federal University of Rio Grande do Norte}\\
Natal, Rio Grande do Norte \\
arthur.romao@ufrn.br}
\and
\IEEEauthorblockN{Bruno Motta de Carvalho}
\IEEEauthorblockA{\textit{Department of Informatics and Applied Mathematics} \\
\textit{Federal University of Rio Grande do Norte}\\
Natal, Rio Grande do Norte \\
bruno@dimap.ufrn.br}
}

\maketitle

\begin{abstract}
Basement relief gravimetry is a key application in geophysics, particularly important for oil exploration and mineral prospecting. It involves solving an inverse geophysical problem, where the parameters of a geological model are inferred from observed data. In this context, the geological model consists of the depths of constant-density prisms representing the basement relief, and the data correspond to the gravitational anomalies caused by these prisms. Inverse geophysical problems are typically ill-posed, as defined by Hadamard, meaning that small perturbations in the data can result in large variations in the solutions. To address this instability, regularization techniques, such as those proposed by Tikhonov, are employed to stabilize the solutions. This study presents a comparative analysis of various regularization techniques applied to the gravimetric inversion problem, including Smoothness Constraints, Total Variation, the Discrete Cosine Transform (DCT), and the Discrete Wavelet Transform (DWT) using Daubechies D4 wavelets. Optimization methods are commonly used in inverse geophysical problems because of their ability to find optimal parameters that minimize the objective function—in this case, the depths of the prisms that best explain the observed gravitational anomalies. The Genetic Algorithm (GA) was selected as the optimization technique. GA is based on Darwinian evolutionary theory, specifically the principle of natural selection, where the fittest individuals in a population are selected to pass on their traits. In optimization, this translates to selecting solutions that most effectively minimize the objective function. The results, evaluated using fit metrics and cumulative error analysis, demonstrate the effectiveness of all the regularization techniques and the Genetic Algorithm. Among the methods tested, the Smoothness constraint was briefly the most effective for the first and second synthetic models. For the third model, which was based on real data, all regularization methods performed equivalently.
\end{abstract}


\section{INTRODUCTION}

The identification of geological features in basement topography is of great importance to the oil industry \citep{telford1990applied}. The study of basement topography in sedimentary basins aids in the search for oil and gas accumulations. A gently undulating basement topography, or one locally marked by a series of small-offset faults, can lead to the trapping of a reservoir rock layer between impermeable rock layers and the basement, forming a stratigraphic trap for oil \citep{santos2013inversao}. Therefore, studying basement topography is crucial for identifying potential oil traps.

Estimating the topography is an inverse problem, as it involves estimating the heights of basement topography prisms from observed gravity data. According to \cite{hadamard1902problemes}, an inverse problem is considered ill-posed if it is characterized by the absence, non-uniqueness, or instability of solutions. The gravimetric inverse problem of estimating basement topography is ill-posed because its solution is not unique and is unstable. To transform it into a well-posed problem, the incorporation of \textit{a priori} geological information is required. This is commonly achieved by adding a regularizing term to the functional \citep{tikhonov1977solutions}.

Gravimetric data interpretation can utilize inversion methods. For example, the inversion method using the smoothness regularizer (SV) \citep{tikhonov1977solutions} can be applied to estimate the topography of sedimentary basins with smooth basements. However, to estimate abrupt discontinuities in the basement topography of sedimentary basins, it is recommended to apply inversion methods that incorporate total variation (TV) constraints, such as those proposed by \cite{martins2011total} and \cite{lima2011total}.

This work addresses the estimation of both smooth and non-smooth solutions for the gravimetric inverse problem of estimating the topography of a sedimentary basin. Regularization methods commonly used in geophysical inversion aim to reconstruct smooth solutions, even though geological structures often exhibit sharp contrasts (discontinuities) in physical properties.

The article by \cite{gholami2010regularization} addresses the regularization of geophysical ill-posed problems, both linear and non-linear, using joint sparsity constraints. We introduce the sparsity constraint using the discrete cosine transform (DCT) and discrete wavelet transforms (DWT) of the Daubechies D4 type \citep{daubechies1992ten} for this inverse problem. These constraints allow the reconstruction of sparse solutions in the DCT and wavelet domains. This approach enables the reconstruction of non-smooth density functions that represent abrupt geological structures. The sparsity constraint is particularly important because it allows for solutions that exhibit sharp edge definitions of anomalous bodies \citep{youzwishen2006edge}.

An inversion algorithm based on the genetic algorithm was implemented in MATLAB to solve the inverse problem using various constraints.

We applied the proposed methodology, using smoothness, total variation, and sparsity constraints, to synthetic data produced by two simulated geological environments with abrupt discontinuities in basement topography. Additionally, we applied the proposed method to real data from the Ponte do Poema, located at the UFPA Belém Campus. The synthetic and real models were based on the works of \cite{lima2009inversao} and \cite{santos2013inversao}, respectively. The objective was to identify which constraint best regularized the problem and provided the best fit to the observed data.

\section{METHODOLOGY}

It is assumed that a sedimentary basin consists of homogeneous basement rocks and heterogeneous sediments. Let $g$ be a set of $N$ gravimetric observations inserted into a Cartesian plane. We assume that the density contrast $\Delta\rho$ between the sediment and the basement is constant and known. From these observations, comparisons between the density of the basement topography $S$ and the sedimentary material are generated.

The $z$ axis is positive vertically downward in the Cartesian coordinate system. The sediment of a basin is approximated by a rectangular matrix of $M$ vertically juxtaposed 2D rectangular prisms, whose tops coincide with the ground surface. The horizontal dimensions of all prisms are constant and known. The depths of the bases of the prisms $p_j$ (with $j = 1, \dots, M$) represent the depths of the basement.

The gravity anomaly $g_i$ at the $i$-th observation point caused by a set of prisms is given by the non-linear relationship:

\begin{equation}
	\label{equacaoAnomaliaGravimetrica1}
	g_i = \sum_{j = 1}^{M} f_i(\Delta \rho_j,p_j,r_{ij}), \;\;\;\;\;\; i = 1, 2, ..., N
\end{equation}

The non-linear function $f_i(\Delta \rho_j, p_j, r_{ij})$ calculates the gravitational anomaly caused by the $j$-th prism at the $i$-th observation point. That is, it solves the forward problem for an individual prism at observation point $i$. Here, $\Delta \rho_j$ is the density contrast of the $j$-th prism, $p_j$ is the depth of the $j$-th prism, and $r_{ij}$ is the horizontal position vector between the $i$-th point and the $j$-th prism \citep{silva2010gravity}.

\begin{figure}[H]
	\centering	\includegraphics[width=0.7\linewidth]{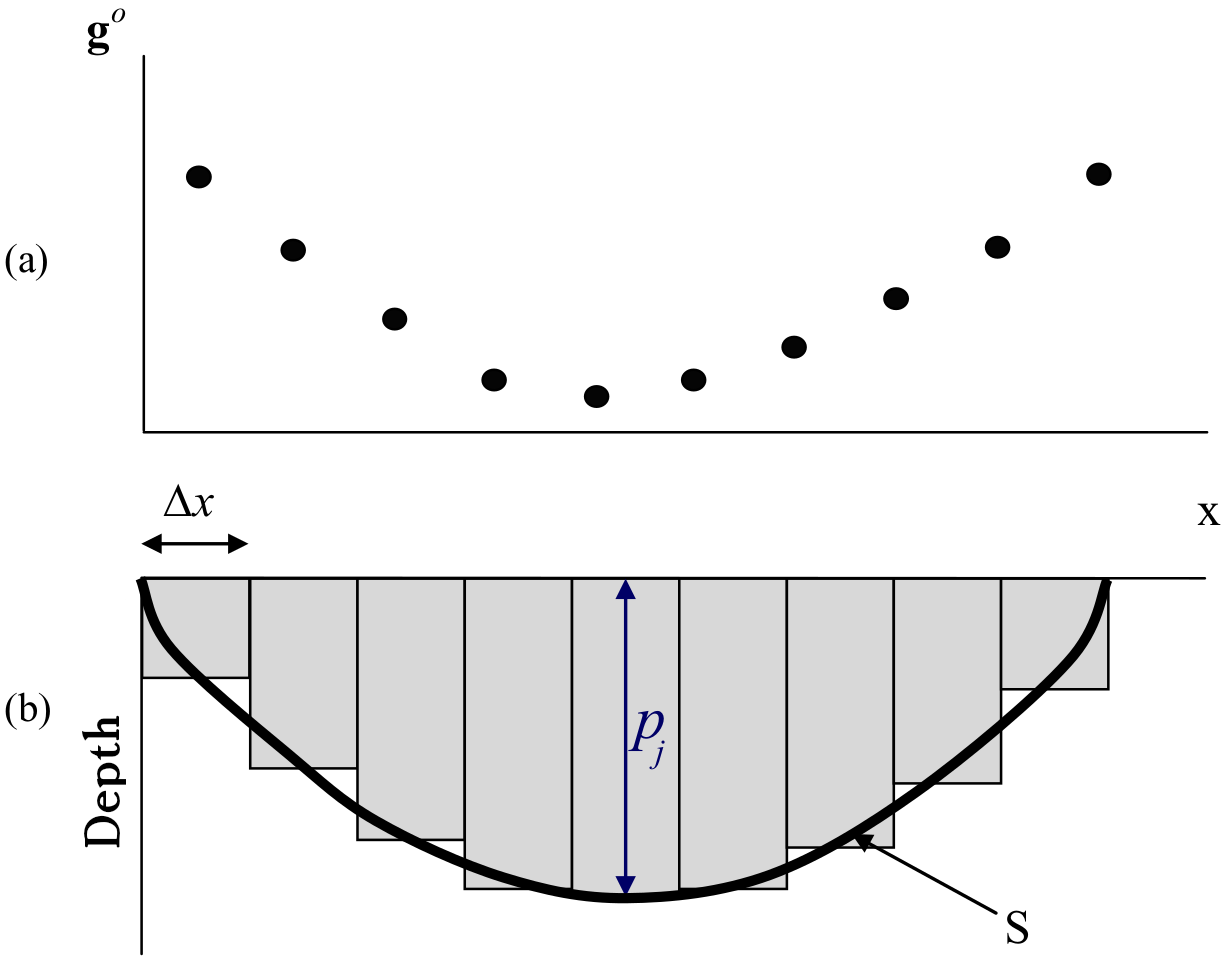}
  \caption{(a) Gravimetric anomaly and (b) sedimentary basin showing basement topography $S$ and an interpretive model formed by a set of $M$ juxtaposed 2D rectangular prisms with height $p_{j}$.\\
\centering\textbf{Source}: \cite{santos2013inversao}}
	\label{fig teste}
\end{figure}

The objective is to estimate $S$ considering, as an interpretive model, a set of $M$ vertical rectangles juxtaposed along the $x$-axis of the Cartesian plane. The depths $p_j$ of the rectangles are the parameters to be estimated. The apex of each rectangle is located at the Earth's surface, collectively forming a flat ground, and all rectangles have the same horizontal distances $\Delta x$ \cite{lima2009inversao}.

\subsection{Inverse Problem}

According to \cite{pitfallsnonlinearinversion}, the geophysical problem of gravimetric inversion can be formulated as the minimization of the functional $\phi$, described by the equation:

\begin{equation}
	\label{funcionalSemLagrange}
	\phi(p) = \dfrac{\lVert g^{obs} - g(p) \rVert ^2_2}{N f_\phi },
\end{equation}

\noindent where $p$ is the depth vector defined as $p = (p_1, p_2, \ldots, p_M)$, $g^{obs} = (g_1^{obs}, g_2^{obs}, \ldots, g_N^{obs})$ represents the set of observed gravity measurements, and $g(p) = (g_1, g_2, \ldots, g_N)$ is the set of calculated gravity values based on Equation \ref{equacaoAnomaliaGravimetrica1}. Here, $N$ denotes the number of observations \citep{valeria,pitfallsnonlinearinversion}, and $f_\phi$ is a normalization factor. It is assumed that $\phi$ is non-linear, continuous, and has continuous first and second derivatives with respect to $p$.

\subsection{Regularization}

The problem of minimizing $\phi(p)$ is ill-posed when finding the vector $p$ and minimizing $\phi(p)$, as the obtained solutions generally exhibit instability. To regularize the problem, a constraint is used, which is \textit{a priori} information added with the aim of regularizing the problem \citep{pitfallsnonlinearinversion}.

Regularization can be understood as the approximation of an ill-posed problem to a family of quasi-solutions \citep{tikhonov1977solutions}. The objective of regularization is to try to circumvent the violation of one or more Hadamard conditions \citep{engl1996regularization}. The objective function added with a constraint then becomes:

\begin{equation}
	\label{funcionalComLagrange}
	\Omega(p) = \phi(p) + \alpha \dfrac{\mathcal{R}(p)}{f_{\mathcal{R}}} ,
\end{equation}

\noindent where $\phi$ is the functional described in Equation \ref{funcionalSemLagrange}, $\alpha$ is the Lagrange multiplier, $\mathcal{R}$ is the constraint that aims to regularize the solution, and $f_{\mathcal{R}}$ is the normalization factor of the constraint.

\subsubsection{Smoothness}
One of the regularizations used in this work is the one adopted by \cite{tikhonov1977solutions}. The objective of the smoothness constraint is to stabilize a solution by favoring a smooth behavior \citep{lima2009inversao,silva2010gravity,barboza2019user}.

Therefore, the inversion process uses a Lagrange multiplier multiplied by the constraint, and this product is added to $\phi$ from Equation \ref{funcionalSemLagrange}. Thus, the inverse geophysical problem is described as:

\begin{equation}
	\label{funcionalComSuavidade}
	\Omega(p) = \phi(p) + \alpha \dfrac{\mathcal{R}_{suav}(p)}{f_{\mathcal{R}_{suav}}} ,
\end{equation}

\noindent where

\begin{equation}
	\label{vinculoSuavidade}
	\mathcal{R}_{suav}(p) = \lVert \mathcal{S}p \rVert ^2 _2,
\end{equation}

\noindent and where $\alpha$ is the Lagrange multiplier, $\mathcal{R}_{suav}$ is the smoothness constraint, $\mathcal{S}$ is the finite difference matrix, and $p$ is the vector of depths of the prisms representing the basin.

\subsubsection{Total Variation}
For the proposed problem, minimizing the squared $L_2$ norm of the depth vector multiplied by the finite difference matrix provides estimates of smooth and undulating topography, whereas minimizing the $L_1$ norm of the same product results in estimates that favor discontinuities \citep{lima2009inversao,santos2013inversao,barboza2019user}.

The total variation (TV) constraint uses the $L_1$ norm, and the inverse geophysical problem can be described as:

\begin{equation}
	\label{funcionalComTV}
	\Theta(p) = \phi(p) + \alpha \dfrac{\mathcal{R}_{TV}(p)}{f_{\mathcal{R}_{TV}}},
\end{equation}

\noindent where
\begin{equation}
	\label{vinculoTV}
	\mathcal{R}_{TV}(p) = \lVert \mathcal{S}p \rVert ^1 _1,
\end{equation}

\noindent where $\alpha$ is the Lagrange multiplier, $\mathcal{R}_{TV}$ is the TV constraint, $\mathcal{S}$ is the finite difference matrix, and $p$ is the vector of depths of the prisms.

\subsubsection{Sparsity}
The sparsity constraint uses the Discrete Cosine Transform (DCT) and the Discrete Wavelet Transform (DWT), and employs the $L_1$ norm to achieve a sparse model \citep{gholami2010regularization,dantas2015resoluccao,barboza2019user}.

A wavelet can be described as a small wave, complementing the classical Fourier decomposition method (which uses large sinusoidal waves), and in essence, it is an oscillation that decays rapidly \citep{sifuzzaman2009application}. The use of DCT results in decorrelation of data, removing redundancy among neighboring data points \citep{khayam2003discrete}.

Both DCT and DWT constraints use the $L_1$ norm, resulting in a sparse model in inversion. The inverse geophysical problem with sparsity constraint is defined as:

\begin{equation}
	\label{funcionalComEsparsidade}
	\Psi(p) = \phi(p) + \alpha \dfrac{\mathcal{R}_{esp}(p)}{f_{\mathcal{R}_{esp}}} ,
\end{equation}

\noindent where

\begin{equation}
	\label{vinculoEsp}
	\mathcal{R}_{esp}(p) = \lVert \mathcal{E}p \rVert ^1 _1,
\end{equation}

\noindent where $\alpha$ is the Lagrange multiplier, $\mathcal{R}_{esp}$ is the sparsity constraint, $p$ is the vector of depths of the basin, and $\mathcal{E}$ is the matrix of the transform, such as the Discrete Wavelet Transform (DWT) or the Discrete Cosine Transform (DCT).

\subsection{Genetic Algorithms}
Genetic Algorithms are search algorithms based on mechanisms of natural selection and genetics \citep{holland1975adaptation}. The purpose of the algorithm is to indicate a potential solution to a specific problem using a data structure similar to chromosomes and applying recombination of these structures to preserve critical information. Additionally, genetic algorithms are often seen as metaheuristic global optimization techniques.

The goal of the algorithm is to iteratively generate increasingly evolved populations through evolutionary rules or genetic operations in pursuit of an optimal result. A population becomes more evolved when its individuals are better adapted to the environmental conditions. In solving the inverse problem, for instance, the smaller the value of the function for individuals in the population, the better adapted they are, implying a more evolved population. The initial population is generated randomly or based on information from the problem to be optimized \citep{golberg1989genetic}.

Initially, an initial population is generated with a set of random chromosomes. Each chromosome represents a possible solution, which is then evaluated to assign a score based on desired characteristics. The best solutions are selected (elitism), and new chromosomes are generated through crossover between chromosomes from the previous iteration and/or mutations. This process is repeated until the algorithm's stopping criterion is met. If we define $pop(t)$ as a population of chromosomes at generation $t$, the standard pseudocode of the genetic algorithm is outlined in Algorithm \ref{alg:genetic} as described by \cite{holland1975adaptation}.

\begin{algorithm}
	\caption{Standard Genetic Algorithm}
	\begin{algorithmic} 
		\STATE Initialize $pop(t)$
		\STATE Evaluate $pop(t)$
		\WHILE{stopping criterion not satisfied}
			\STATE $t \leftarrow t + 1$
			\STATE Select $pop(t)$ from $pop(t-1)$
			\STATE Apply crossover in $pop(t)$
			\STATE Apply mutation in $pop(t)$
			\STATE Apply elitism in $pop(t)$ from $pop(t-1)$
			\STATE Evaluate $pop(t)$
		\ENDWHILE
	\end{algorithmic}
    \label{alg:genetic}
\end{algorithm}

In this study, the population size was 1,000, the maximum number of iterations was 1,000, the tournament size was 3, the mutation rate was $1\%$, and the mutation magnitude was $10\%$. The crossover rate was $75\%$, the lower bound was a vector of zeros with the size corresponding to the number of parameters to be estimated, and the upper bound was a vector with the same number of parameters, but with a fixed value of 4, based on a maximum depth estimate of the synthetic and real models studied. These parameters encompass a more comprehensive version of the Genetic Algorithm, as detailed by the authors \cite{holland1975adaptation, goldberg1987genetic, haupt2004practical, man1999genetic, michalewicz1994evolution}.

\section{RESULTS AND DISCUSSION}

In this section, two synthetic models simulating gravimetric anomalies in sedimentary basins are presented. Both models are two-dimensional, with constant density contrast, geologically smooth, but featuring significant discontinuities that are to be estimated through inversion.

Gravimetric data from these synthetic models were generated and had 5\% pseudo-random Gaussian noise added to simulate the variations typically observed during field measurements.

Additionally, real data collected from the Ponte do Poema, located at the Federal University of Pará, Belém Campus, were used to validate the inversion methods applied in this study. The Ponte do Poema was chosen for its strategic location and structural characteristics, which provide stable conditions for gravimetric measurements. This real data offers a practical context for evaluating the performance of the proposed inversion techniques, providing insight into how these methods function under real-world conditions.

For each model, 10 inversions were conducted for each of the four constraints addressed in this study. From the inversion results, the average solution for each constraint was calculated, resulting in different mean models with distinct characteristics depending on the applied constraint. The Lagrange multipliers ($\alpha$) for each constraint were determined using the L-curve method \citep{hansen1999curve}.

Moreover, figures showing the convergence of functionals from the inversions with each proposed constraint are presented. It is important to note that each figure represents an example of a single inversion for each constraint, aiming to illustrate the behavior of the functionals. However, the isolated functionals do not represent the optimal solution, as they do not fully integrate the constraints. Nonetheless, they exhibit values that are close to one another, considering the scale of the graphs.

\subsection{Isolated Graben}

For this model, a constant density contrast of -0.25 g/cm$^{3}$ was established. The horizontal extent of the model is 60 km, using 60 prisms, and the maximum depth reaches 1.85 km.

\begin{figure}[H]
\centering
%\captionsetup{justification=centering}
\includegraphics[width=1\linewidth]{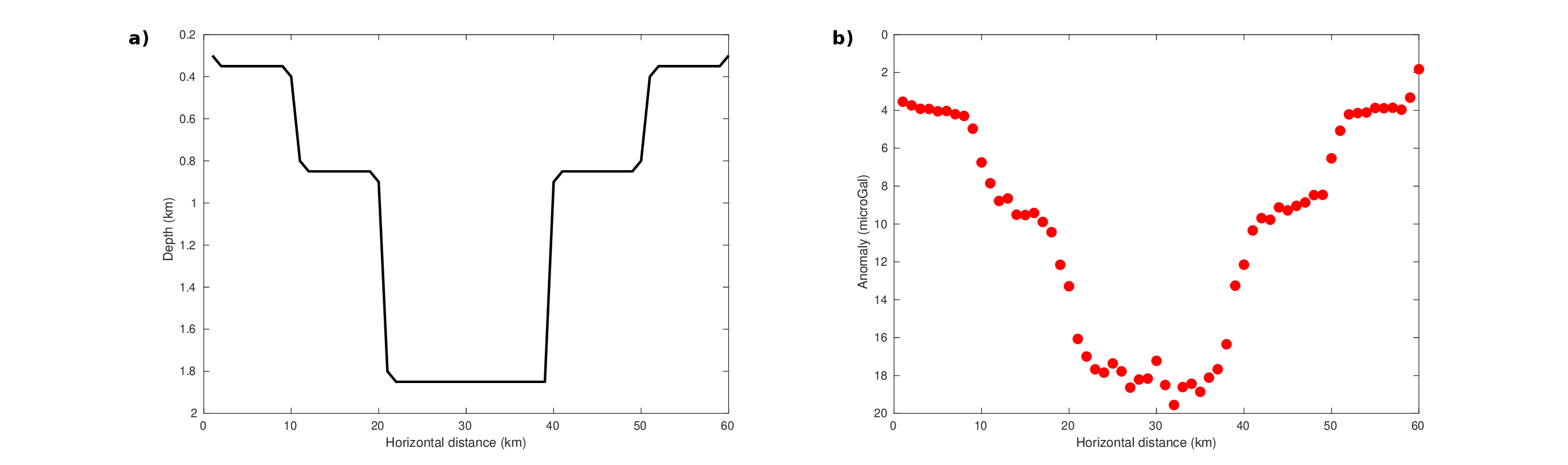}
\caption{(a) True synthetic model of isolated graben and (b) observed gravimetric data set from the model in question.}
\label{subplotGIGrav}
\end{figure}

Figure \ref{subplotGIGrav} synthetically presents a model of a sedimentary basin with an isolated graben at the center of the measurements, as described by \cite{lima2009inversao}. Gravimetric measurements were generated based on this synthetic model.

\begin{figure}[H]
\centering
%\captionsetup{justification=centering}
\includegraphics[width=1\linewidth]{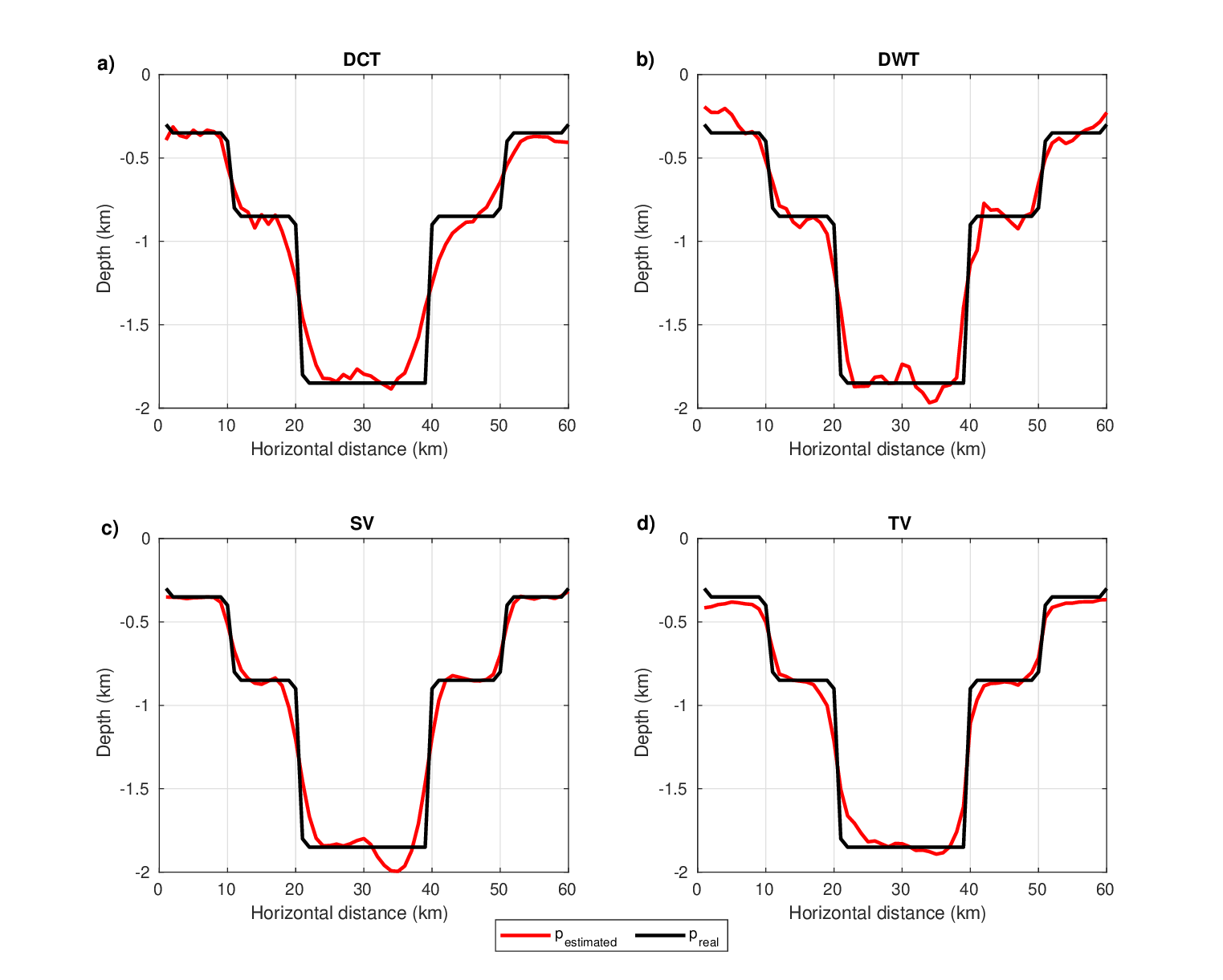}
\caption{Result of the average of 10 inversions of the isolated graben model using constraint (a) DCT, (b) DWT, (c) SV, and (d) VT.}
\label{subplotsGI}
\end{figure}

The Figure \ref{subplotsGI} presents the result of the average of 10 inversions estimated based on observed gravimetric data, along with the true synthetic model for comparison in each constraint. The Lagrange multipliers were $\alpha=8 \times 10^{-1}$ for DCT, $\alpha=8 \times 10^{-1}$ for TV, $\alpha=9 \times 10^{-1}$ for DWT, and $\alpha=2 \times 10^{0}$ for SV. The normalization factors are shown in Table \ref{tbl_factor_normalization_gi}.

\begin{table}[!htpb]
\begin{tabular}{|c|c|c|l|}
\hline
\textbf{$\Omega$}                                                                                     & \textbf{$\Theta$}                                                                                   & \textbf{$\Psi_{DCT}$}                                                                                & \multicolumn{1}{c|}{\textbf{$\Psi_{DWT}$}}                                                             \\ \hline
\begin{tabular}[c]{@{}c@{}}$f_{\phi} = 1.5$\\ $f_{\mathcal{R}_{suav}} = 1$\end{tabular} & \begin{tabular}[c]{@{}c@{}}$f_{\phi} = 1.5$\\ $f_{\mathcal{R}_{TV}} = 1$\end{tabular} & \begin{tabular}[c]{@{}c@{}}$f_{\phi} = 1$\\ $f_{\mathcal{R}_{DCT}} = 1.1$\end{tabular} & \begin{tabular}[c]{@{}l@{}}$f_{\phi} = 0.95$\\ $f_{\mathcal{R}_{DWT}}= 1.2$\end{tabular} \\ \hline
\end{tabular}
\label{tbl_factor_normalization_gi}
\caption{Normalization factors for the Isolated Graben Model.}
\end{table}

\begin{figure}[H]
\centering
%\captionsetup{justification=centering}
\includegraphics[width=0.8\linewidth]{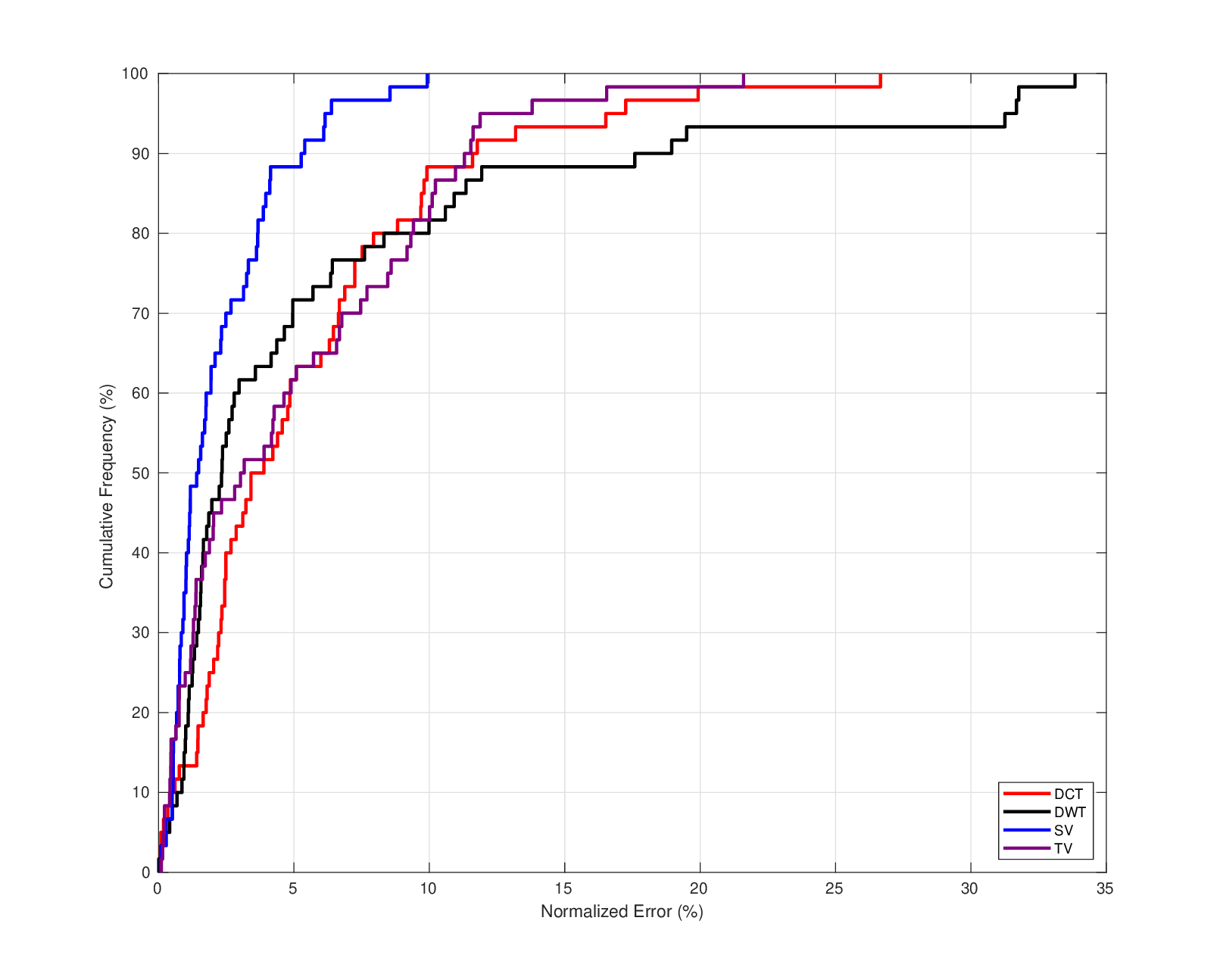}
\caption{Relative error between the data from the true isolated graben model and the average of 10 inversions in each addressed constraint.}
\label{errosRelativosGI}
\end{figure}

The estimated parameters, as shown in Figure \ref{subplotsGI}, indicate a good fit among the constraints. For a more detailed analysis, the relative error between the data estimated by the average of inversions and the data from the true synthetic model was calculated, as shown in Figure \ref{errosRelativosGI}.

Figure \ref{errosRelativosGI} shows that the best results, based on the average of inversions, were achieved with the DCT constraint, where 89\% of the data contributed to approximately 10\% of the error. The TV constraint produced an average inversion error of about 12\%, also associated with 89\% of the data. For the DWT constraint, the average error was around 13\%, with 89\% of the data considered. The best overall result came from the SV constraint, where 89\% of the data accounted for only about 4\% of the error.

\begin{figure}[H]
\centering
%\captionsetup{justification=centering}
\includegraphics[width=0.8\linewidth]{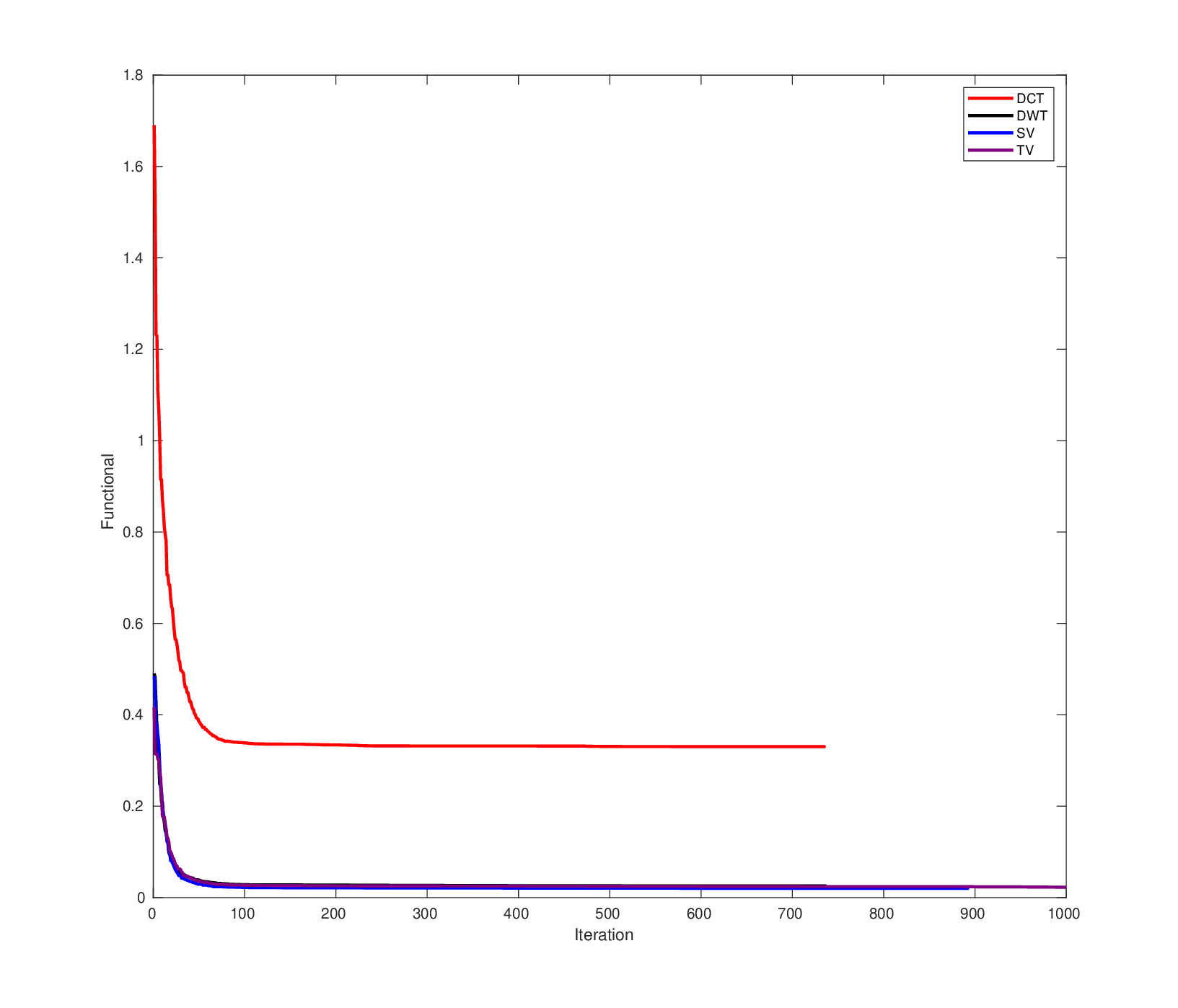}
\caption{Convergence curve of the functional with the constraints used in the inversion of gravimetric data from the isolated graben model.}
\label{funcionalGI}
\end{figure}

Figure \ref{funcionalGI} illustrates the inversion functional, which graphically represents the difference between observed data and estimated data, considering the constraint used.

Figure \ref{funcionalGI} shows the convergence of the functional discussed in Equation \ref{funcionalComLagrange}. It can be observed that there was good and equivalent convergence of functionals among the constraints.

\subsection{Two Sub-basins}

For this model, a constant density contrast of -0.25 g/cm$^{3}$ was established. The horizontal extent of the model is 80 km, using 80 prisms, and the maximum depth reaches 3.5 km.

\begin{figure}[H]
\centering
%\captionsetup{justification=centering}
\includegraphics[width=1\linewidth]{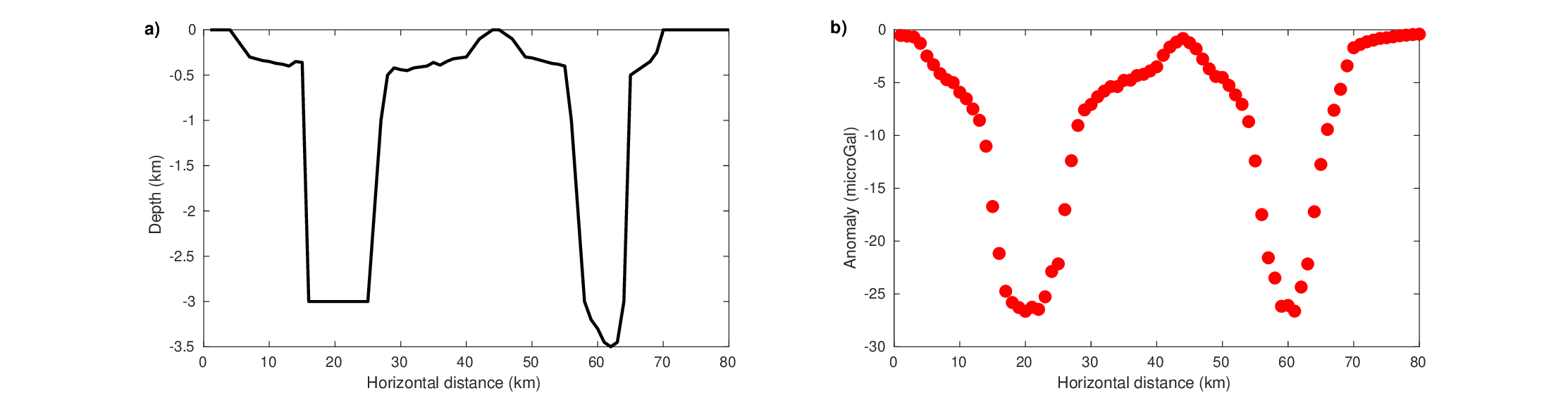}
\caption{(a) True synthetic model of two sub-basins and (b) observed gravimetric data set from the model in question. In (b), it is noticeable that the depocenter of the left basin has greater curvature than the depocenter of the right basin.}
\label{subplotDBGrav}
\end{figure}

Figure \ref{subplotDBGrav} synthetically presents a model of a sedimentary basin with two consecutive and closely spaced sub-basins, as described by \cite{lima2009inversao}. Gravimetric measurements were generated based on this synthetic model.

\begin{figure}[H]
\centering
%\captionsetup{justification=centering}
\includegraphics[width=1\linewidth]{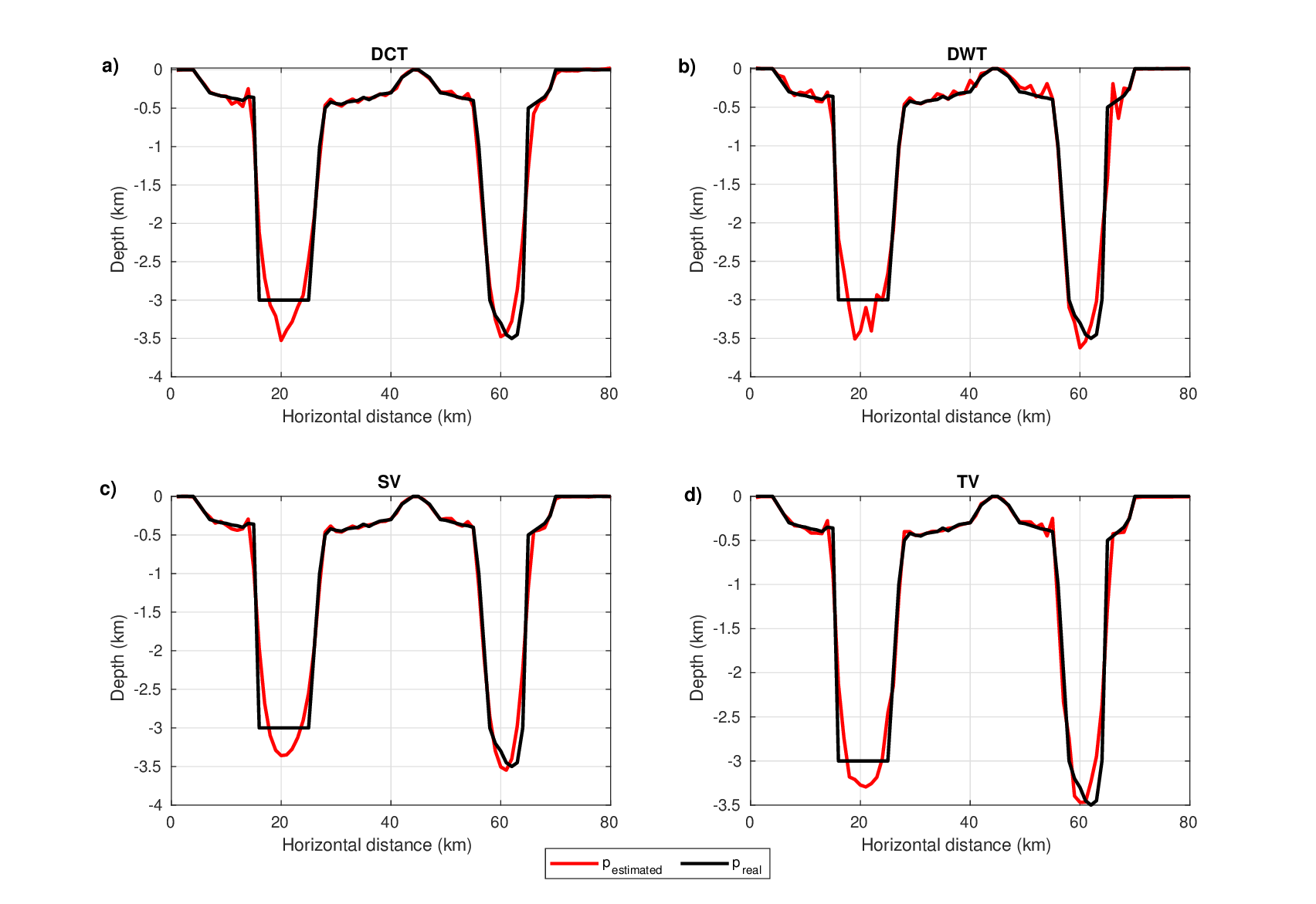}
\caption{Result of averaging 10 inversions of the two sub-basin model using (a) DCT, (b) DWT, (c) SV, and (d) TV constraints.}
\label{subplotDB}
\end{figure}

Figure \ref{subplotDB} presents the result of averaging 10 inversions estimated based on the observed gravimetric data, along with the true synthetic model for comparison under each constraint. The Lagrange multipliers were $\alpha=1 \times 10^{-1}$ for DCT, $\alpha=1 \times 10^{-1}$ for TV, $\alpha=3 \times 10^{-2}$ for DWT, and $\alpha=25 \times 10^{-2}$ for SV. The normalization factors are shown in Table \ref{tbl_factor_normalization_db}.

\begin{table}[!htpb]
\begin{tabular}{|c|c|c|l|}
\hline
\textbf{$\Omega$}                                                                                     & \textbf{$\Theta$}                                                                                   & \textbf{$\Psi_{DCT}$}                                                                                   & \multicolumn{1}{c|}{\textbf{$\Psi_{DWT}$}}                                                              \\ \hline
\begin{tabular}[c]{@{}c@{}}$f_{\phi} = 1.6$\\ $f_{\mathcal{R}_{suav}} = 1$\end{tabular} & \begin{tabular}[c]{@{}c@{}}$f_{\phi} = 1.6$\\ $f_{\mathcal{R}_{TV}} = 1$\end{tabular} & \begin{tabular}[c]{@{}c@{}}$f_{\phi} = 2.1$\\ $f_{\mathcal{R}_{DCT}} = 1.25$\end{tabular} & \begin{tabular}[c]{@{}l@{}}$f_{\phi} = 1.26$\\ $f_{\mathcal{R}_{DWT}}= 1.15$\end{tabular} \\ \hline
\end{tabular}
\label{tbl_factor_normalization_db}
\caption{Normalization factors for the Two Sub-Basins Model.}
\end{table}

\begin{figure}[H]
\centering
%\captionsetup{justification=centering}
\includegraphics[width=0.8\linewidth]{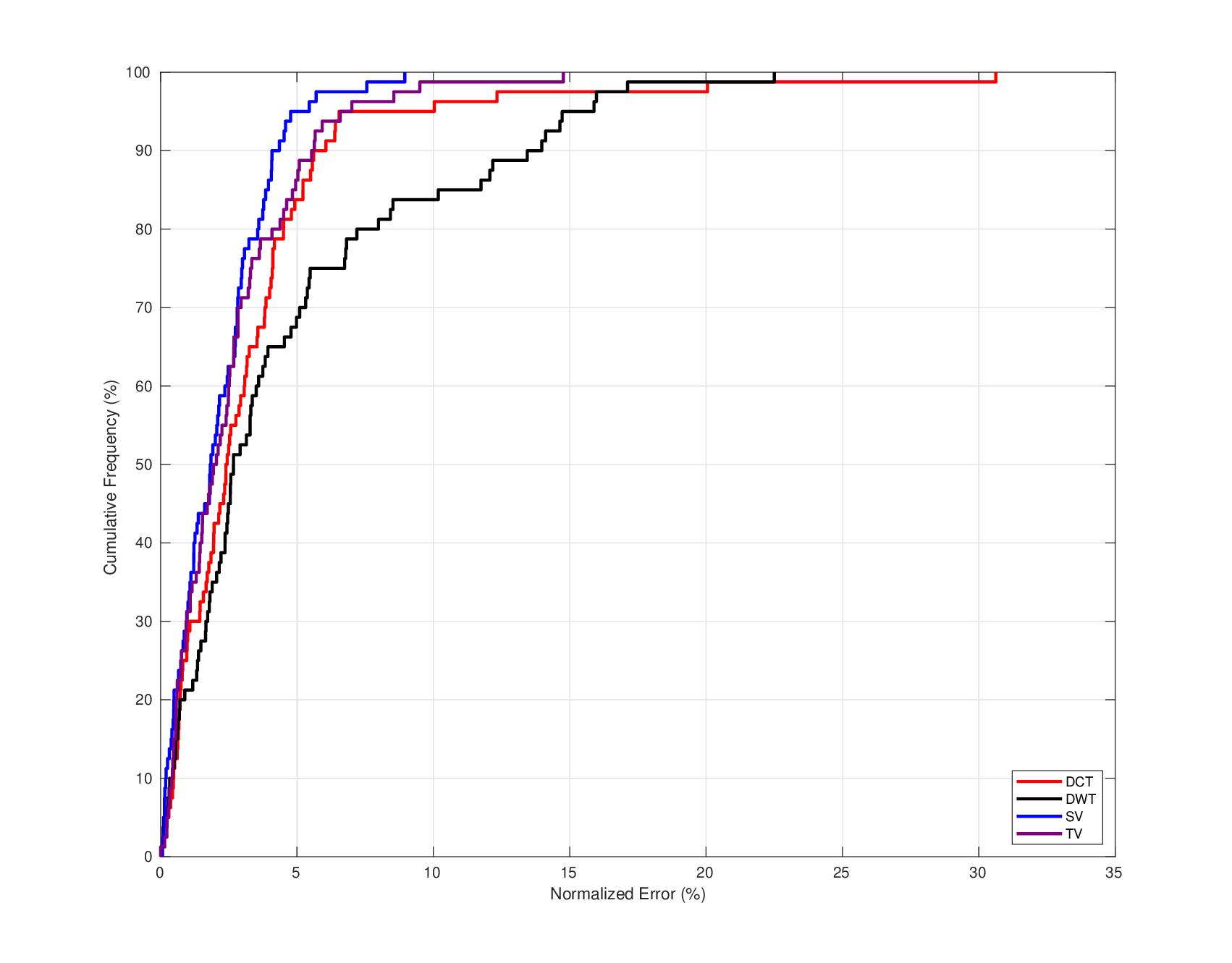}
\caption{Relative error between true two sub-basin model data and the average of 10 inversions under each constraint.}
\label{erroRelativoDB}
\end{figure}

It is observable from Figure \ref{subplotDB} that the constraints were successful in regularization. For a more detailed analysis of the results, Figure \ref{erroRelativoDB} shows the relative error between the estimated data from each inversion and the data from the true synthetic model.

Figure \ref{erroRelativoDB} shows that the best result from the average of inversions was achieved with the SV constraint, where 94\% of the data accounted for approximately 4\% of the error. Inversions using the TV and DCT constraints resulted in around 5.3\% error, also associated with 94\% of the data. Finally, the DWT constraint produced approximately 17\% error, associated with 90\% of the data.

\begin{figure}[H]
\centering
%\captionsetup{justification=centering}
\includegraphics[width=0.8\linewidth]{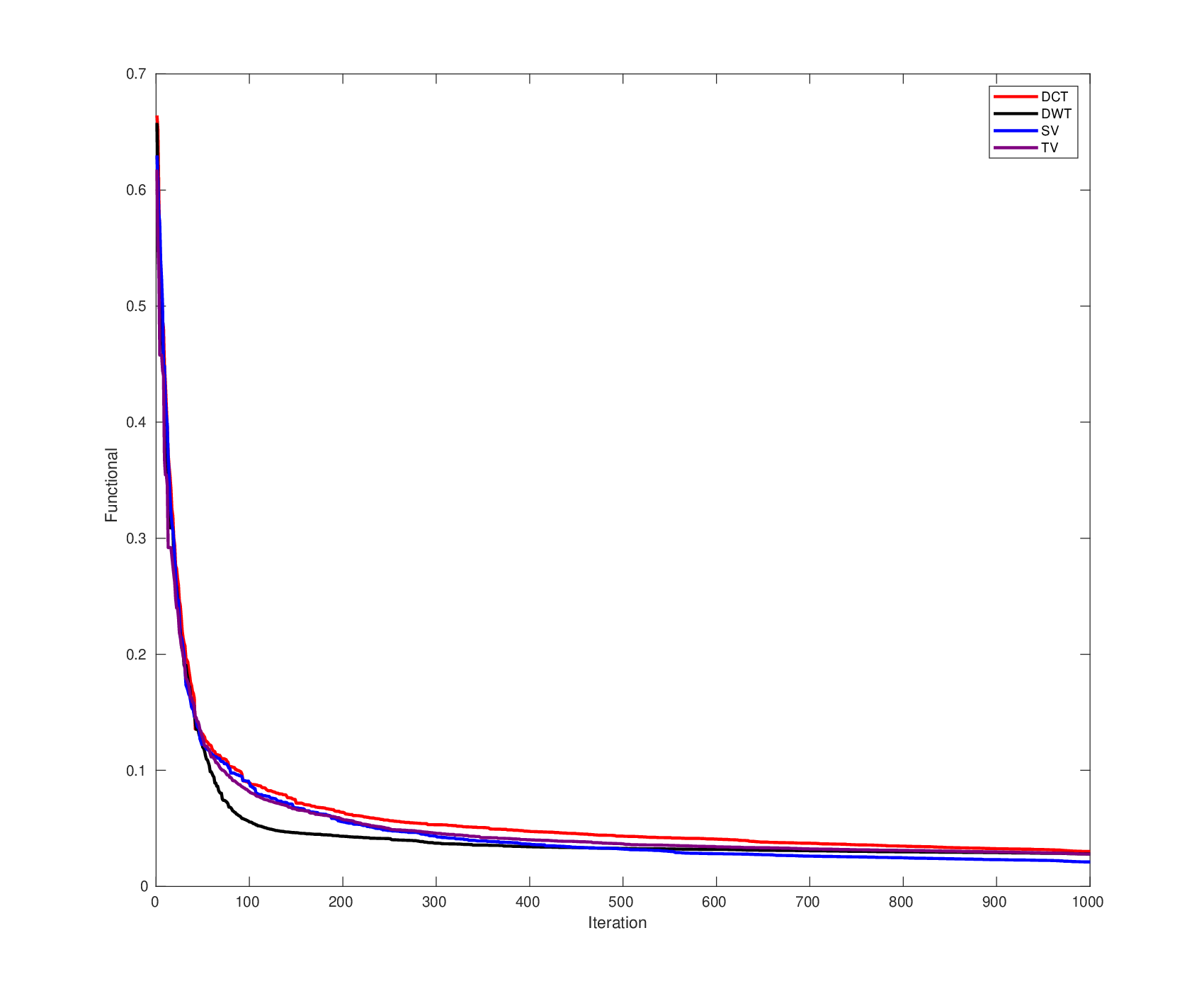}
\caption{Convergence curve of the functional with the constraints used in the inversion of gravimetric data for the two sub-basin model.}
\label{funcionalDB}
\end{figure}

Figure \ref{funcionalDB} illustrates the functional of the inversions, graphically representing the difference between observed and estimated data, considering the constraint used.

Figure \ref{funcionalDB} shows the convergence of the functional discussed in Equation \ref{funcionalComLagrange}. It is evident that there was good and equivalent convergence of the functionals.

\subsection{Bridge of the Poema}

According to the works of \cite{lima2009inversao} and \cite{santos2013inversao}, the methodologies of gravimetric inversion were applied in this study to the observed data at Bridge of the Poema, located at the Belém Campus of the Federal University of Pará.

\begin{figure}[H]
	\centering
	%\captionsetup{justification=centering}
	\includegraphics[scale=0.19]{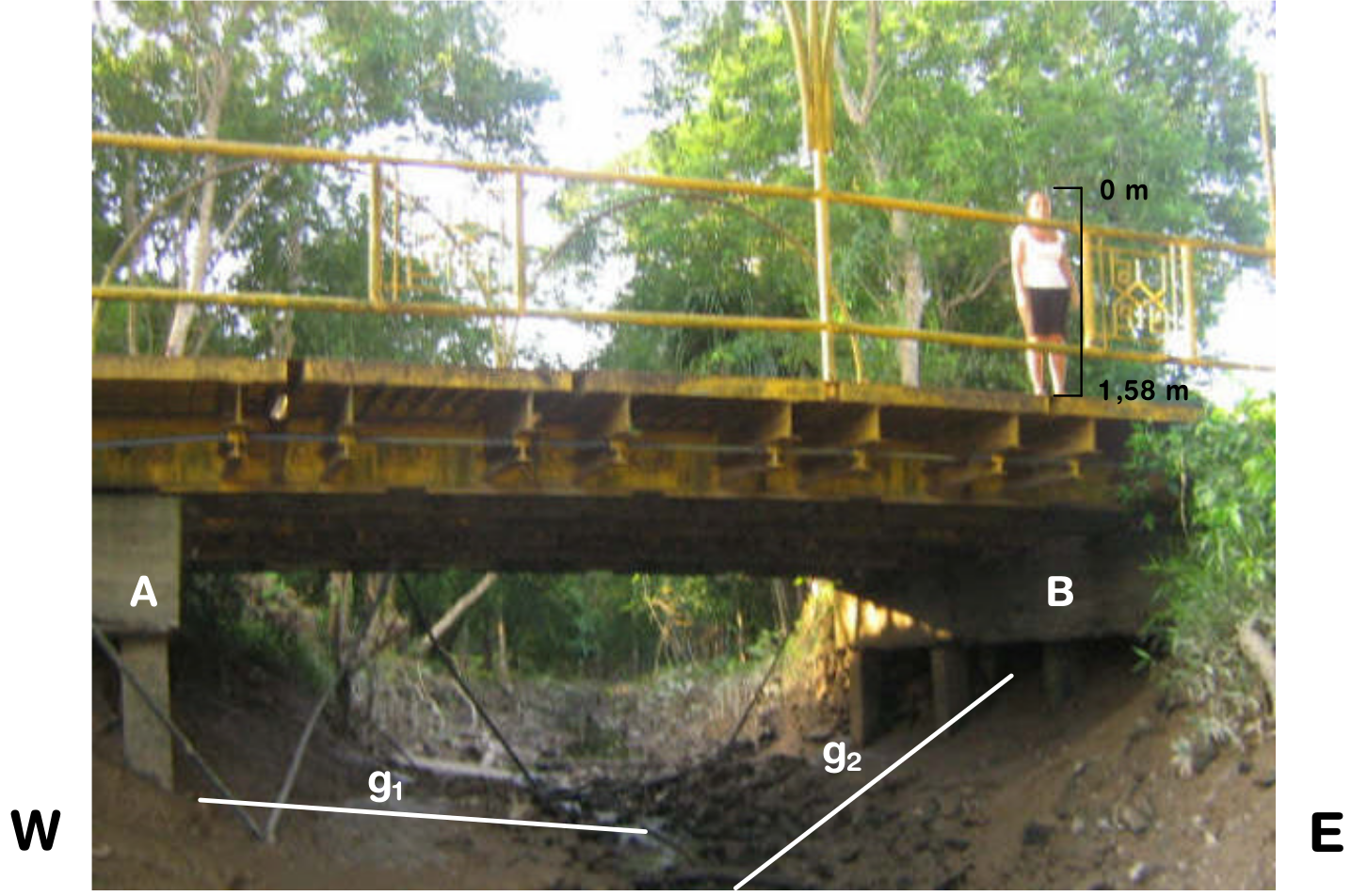}
  \caption{Photograph with W-E gravimetric profile of Bridge of the Poema. The west side has a sharper discontinuity compared to the east side.\\
	\textbf{Source:} \cite{oliveira2007inversao}.}
	\label{fotoPoema}
\end{figure}

Figure \ref{fotoPoema} shows a photograph of the relief of Bridge of the Poema, with the topographic gradients $g_1$ and $g_2$, and the reinforcement piles $A$ and $B$, which are also considered sources of the gravimetric anomaly \citep{lima2009inversao}.

For this interpretative model, a constant density contrast of -2.3 g/cm$^{3}$ was adopted. The horizontal extent of the model is 21.5 m, using 45 prisms. The maximum depth of the relief, estimated from the photograph in Figure \ref{fotoPoema}, is 3.3 m, as described by \cite{oliveira2007inversao}.

\begin{figure}[H]
	\centering
	%\captionsetup{justification=centering}
	\includegraphics[width=1\linewidth]{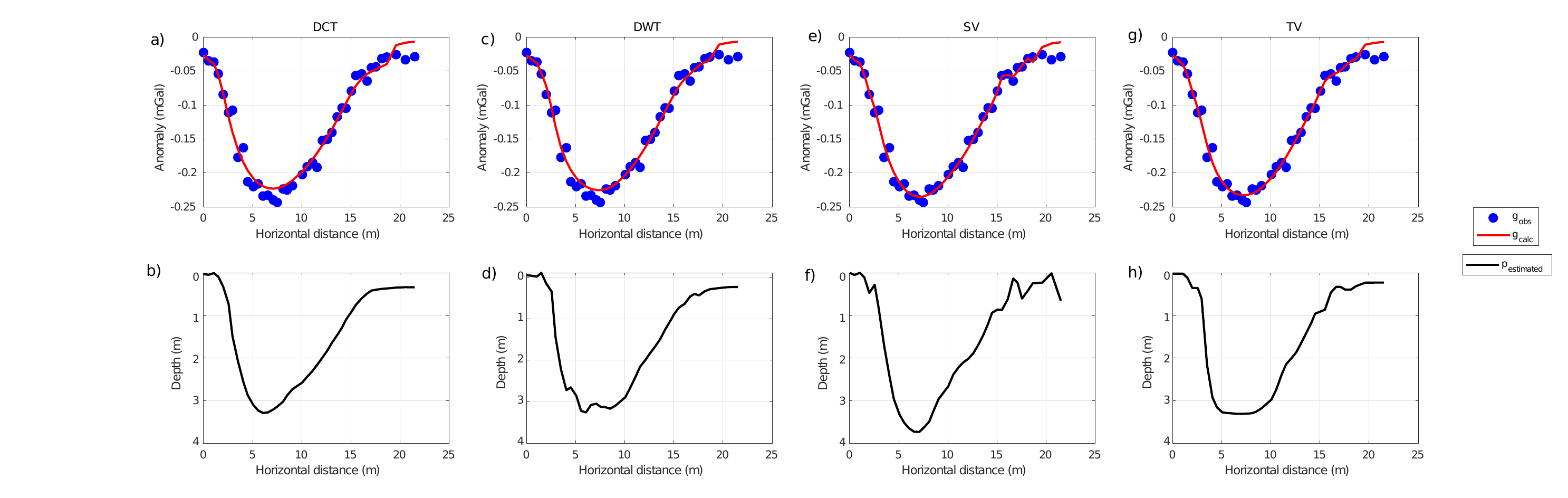}
  \caption{Result of averaging 10 inversions with constraints. Set of gravimetric observations (blue dots) and calculated (red line) from DCT, DWT, SV, and TV constraints, as in (a), (c), (e), and (g), respectively. Inversely estimated model by DCT, DWT, SV, and TV constraints, as in (b), (d), (f), and (h), respectively.}
	\label{subplotPoema}
\end{figure}

Figure \ref{subplotPoema} presents the result of averaging 10 inversions estimated based on gravimetric data using all constraints. The Lagrange multiplier had a value of $\alpha = 2 \times 10^{-3}$ for TV, $\alpha = 1 \times 10^{-3}$ for SV, $\alpha = 6 \times 10^{-3}$ for DCT and DWT. The normalization factors are shown in Table \ref{tbl_factor_normalization_pp}.

\begin{table}[!htpb]
\begin{tabular}{|c|c|c|l|}
\hline
\textbf{$\Omega$}                                                                                   & \textbf{$\Theta$}                                                                                   & \textbf{$\Psi_{DCT}$}                                                                              & \multicolumn{1}{c|}{\textbf{$\Psi_{DWT}$}}                                                        \\ \hline
\begin{tabular}[c]{@{}c@{}}$f_{\phi} = 1$\\ $f_{\mathcal{R}_{suav}} = 1$\end{tabular} & \begin{tabular}[c]{@{}c@{}}$f_{\phi} = 1.5$\\ $f_{\mathcal{R}_{TV}} = 1$\end{tabular} & \begin{tabular}[c]{@{}c@{}}$f_{\phi} = 1$\\ $f_{\mathcal{R}_{DCT}} = 1$\end{tabular} & \begin{tabular}[c]{@{}l@{}}$f_{\phi} = 1$\\ $f_{\mathcal{R}_{DWT}}= 1$\end{tabular} \\ \hline
\end{tabular}
\label{tbl_factor_normalization_pp}
\caption{Normalization factors for the Poema Bridge Model.}
\end{table}

It is notable that all constraints were effective in regularization, and there was a good fit between observed and calculated data, resulting in a model estimate that is quite similar across the adopted constraints. For further evaluation of the best results, refer to Figure \ref{erroRelativoPoema}, which shows the relative error between the average calculated gravities from each inversion and the observed gravimetric data:

\begin{figure}[H]
	\centering
	%\captionsetup{justification=centering}
	\includegraphics[width=0.9\linewidth]{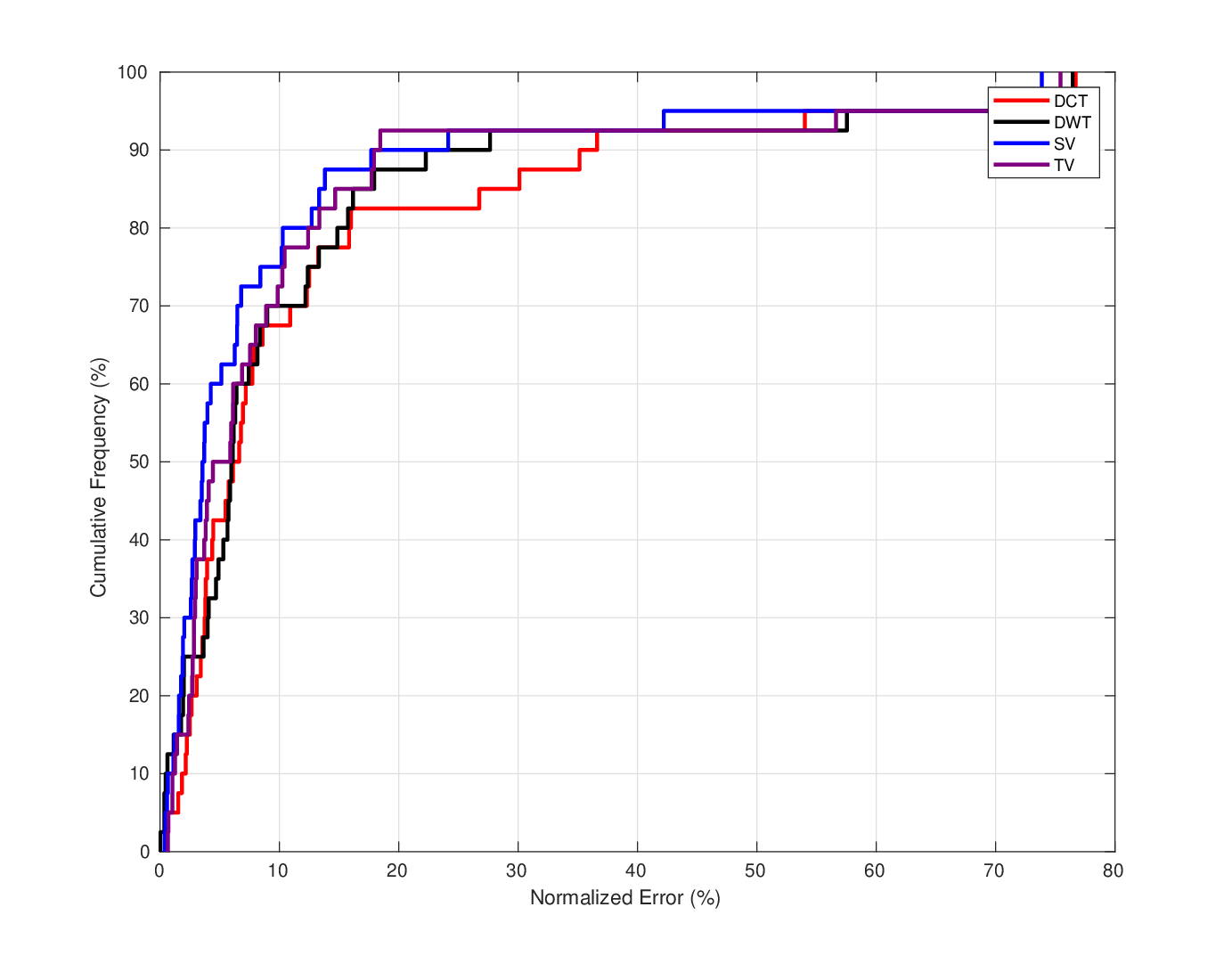}
	\caption{Relative error between the gravimetric data of the Bridge of the Poema model and the average calculated gravities from 10 inversions with each addressed constraint.}
	\label{erroRelativoPoema}
\end{figure}

In Figure \ref{erroRelativoPoema}, it can be observed that for inversions using TV and SV constraints, approximately 90\% of the data accounted for around 17\% error in relation to the observed gravities. For DWT and DCT, 80\% of the data accounted for 14\% error.

\begin{figure}[H]
	\centering
	%\captionsetup{justification=centering}
	\includegraphics[width=0.9\linewidth]{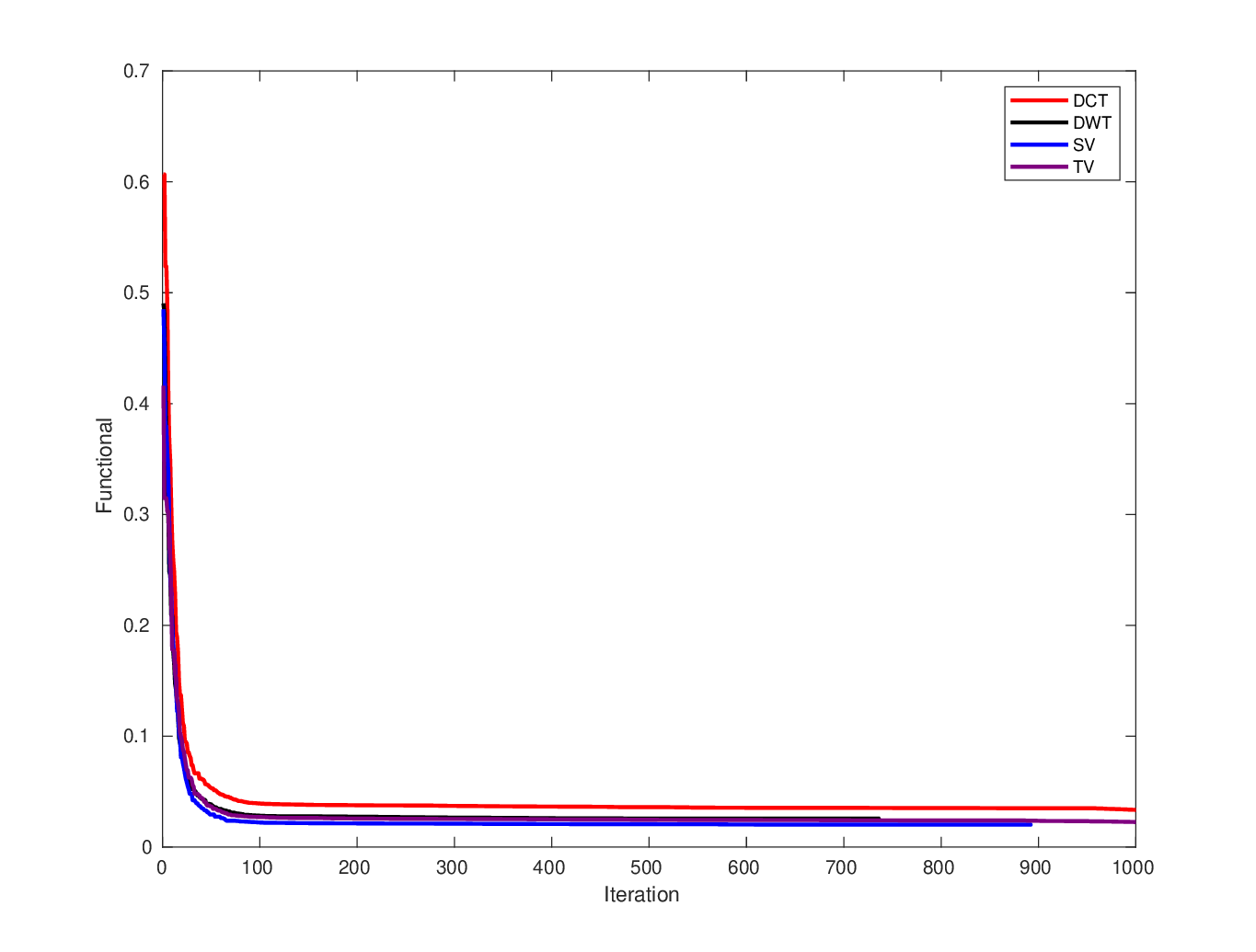}
	\caption{Convergence curve of the functional with the constraints used in the inversion of gravimetric data from the Bridge of the Poema model.}
	\label{funcionalPoema}
\end{figure}

Figure \ref{funcionalPoema} displays the functionals of the inversions, showing excellent and equivalent convergence among the functionals with the addressed constraints.

The limited number of parameters (approximately 100) may have constrained the efficiency of the sparsity enforcement, as models with fewer parameters tend to be less sparse in the DCT or DWT domain.

\section{CONCLUSION}

This study demonstrates the effectiveness of various regularization methods and optimization techniques in addressing the gravimetric inversion problem for 2D basement relief. Among the analyzed methods, the Smoothness regularization technique proved to be the most effective in two synthetic models, briefly reducing the error between the observed and calculated data. However, in the Poema Bridge model, all constraints performed equally well, with a slight advantage over the sparse constraints. Although the Discrete Cosine Transform (DCT) and Discrete Wavelet Transform (DWT) with Daubechies D4 wavelets were less effective in this particular case, their sparse regularization capabilities may perform better in problems involving a larger number of parameters. In this study, the relatively small number of parameters (around 100) may have limited the effectiveness of these techniques. The Genetic Algorithm used for optimization also performed well, efficiently minimizing the objective function and confirming its suitability for solving inverse geophysical problems. Future research could explore the performance of DCT and DWT in more complex models with higher parameter counts, where their ability to impose sparsity may lead to improved outcomes. In gravity inversion, the incorporation of regularization constraints is crucial to obtain interpretable models, as their absence can lead to instability. In this study, all constraints were found to be optimal and equivalent, each offering distinct characteristics. For future work, it is recommended to explore multi-objective algorithms and inversion techniques in 3D models. This could significantly advance research into the application of sparsity constraints in gravity inverse problems and further validate the effectiveness of commonly used regularizers.

\section*{Acknowledgements}
This study was financed in part by the Coordenação de Aperfeiçoamento de Pessoal de Nível Superior - Brasil (CAPES) - Finance Code 001.

\bibliography{reference}

\end{document}